\documentclass{llncs}
\usepackage[top=1.5in, bottom=1.5in,right=1.5in,left=1.5in]{geometry}                
\usepackage{graphicx}
\usepackage{amssymb}
\usepackage{amsmath}
\usepackage{epstopdf}
\usepackage{xspace}
\usepackage{relsize}
\usepackage{bbold}
\usepackage[numbers]{natbib}
\usepackage[colorinlistoftodos,textsize=small]{todonotes}

\usepackage{caption}

\makeatletter
\renewcommand\bibsection%
{
  \section*{\refname
    \@mkboth{\MakeUppercase{\refname}}{\MakeUppercase{\refname}}}
}
\makeatother

\usepackage[applemac]{inputenc} 

\newcommand{\RNAmutants}{\texttt{RNAmutants}\xspace}
\newcommand{\RNApyro}{\texttt{RNApyro}\xspace}

\newcommand{\Z}[3]{\mathcal{Z}_{#1, #2}^{#3}}
\newcommand{\Y}[3]{\mathcal{Y}_{#1, #2}^{#3}}
\newcommand{\B}{\mathcal{B}}
\newcommand{\Kron}{\delta}

\newcommand{\ub}{\bullet}
\newcommand{\op}{\text{\tt(}}
\newcommand{\cp}{\text{\tt )}}

\newcommand{\Struct}{S}

\newcommand{\PE}[1]{E(#1)}
\newcommand{\EI}{\text{EI}}
\newcommand{\ES}{\text{ES}}
\newcommand{\ISO}{\text{ISO}}

\newcommand{\EBP}[3]{E^{(#1)}_{{#2}\to{#3}}}

\newcommand{\Ab}{{\sf{A}}}
\newcommand{\Cb}{{\sf{C}}}
\newcommand{\Gb}{{\sf{G}}}
\newcommand{\Ub}{{\sf{U}}}

\newcommand{\ShowTODO}[1]{{#1}}
\renewcommand{\ShowTODO}[1]{}

\newcommand{\TODO}[2]{\ShowTODO{\todo[inline, linecolor=#1, backgroundcolor=#1!60!white,bordercolor=#1]{#2}}}

\newcommand{\TODOTous}[1]{\TODO{orange}{{\bf TODO Tous :} #1}}
\newcommand{\TODOJerome}[1]{\TODO{blue!80!white}{{\bf TODO Jerome :} #1}}
\newcommand{\TODOYann}[1]{\TODO{gray}{{\bf TODO Yann :} #1}}
\newcommand{\TODOVlad}[1]{\TODO{green!60!black}{{\bf TODO Vlad :} #1}}

\newcommand{\SpaceCheating}{\vspace{-0em}}
\newcommand{\ScaleDP}{.55}

\colorlet{StressColor}{red!60!black}

\title{Using structural and evolutionary information to detect and correct pyrosequencing errors in non-coding RNAs}
\author{Vladimir Reinharz\inst{1}, Yann Ponty\inst{2}$^{*}$ \and J\'er\^{o}me Waldisp\"{u}hl\inst{1}$^*$}
\date{}
\institute{School of Computer Science, McGill University, Montreal, Canada.
	\and  Laboratoire d'informatique, \'Ecole Polytechnique, Palaiseau, France.
	 \\\email{jeromew@cs.mcgill.ca}, \email{yann.ponty@lix.polytechnique.fr}}

\usepackage{subcaption}
\begin{document}

\ShowTODO{\setcounter{tocdepth}{1}
\listoftodos} 
\maketitle
\begin{abstract}
Analysis of the sequence-structure relationship in RNA molecules are essential to evolutionary studies but also to 
concrete applications such as error-correction methodologies in sequencing technologies. The prohibitive sizes of the
mutational and conformational landscapes combined with the volume of data to proceed require efficient algorithms 
to compute sequence-structure properties. More specifically, here we aim to calculate which mutations increase the most the 
likelihood of a sequence to a given structure and RNA family.\\
In this paper, we introduce \RNApyro, an efficient linear-time and space inside-outside algorithm that computes exact mutational
probabilities under secondary structure and evolutionary constraints given as a multiple sequence alignment with a consensus structure.
We develop a scoring scheme combining classical stacking base pair energies to novel isostericity scales, and apply our techniques
to correct point-wise errors in 5s rRNA sequences. Our results suggest that \RNApyro is a promising algorithm to complement existing
tools in the NGS error-correction pipeline. 

\noindent
\textbf{Key words:} RNA, mutations, secondary structure
\end{abstract}


\section{Introduction}
\label{sec:introduction}

\TODOTous{Relire le papier et ajouter references apparues entre temps (ex: en preparant la presentation).}
Ribonucleic acids (RNAs) are found in every living organism, and exhibit a broad range of functions, ranging from catalyzing
chemical reactions, as the RNase P or the group II introns, hybridizing  messenger RNA to regulate gene expressions,
to ribosomal RNA (rRNA) synthesizing proteins. Those functions  require specific structures,  encoded in their nucleotide
sequence. Although the functions, and thus the structures, need to be preserved through various organisms, the sequences
can greatly differ from one organism to another. This sequence diversity coupled with the structural conservation is a fundamental
asset for evolutionary studies. To this end, algorithms to analyze the relationship between RNA mutants and structures are required.

For half a century, biological molecules have been studied as a proxy to understand evolution~\cite{Zuckerkandl1965}, and due
to their fundamental functions and remarkably conserved structures, rRNAs have always been a prime candidate for phylogenetic
studies~\cite{Olsen1986, Olsen1993}. In recent years, studies as the \emph{Human Microbiome Project}~\cite{Turnbaugh2007} benefited
of new technologies such as the NGS techniques to sequence as many new organisms as possible and extract an unprecedented flow of new information. 
Nonetheless, these high-throughput techniques typically have high error rates that make their applications to metagenomics (a.k.a. environmental
genomics) studies challenging. For instance, pyrosequencing as implemented by Roche's 454 produces may have an error rate raising up to 10\%.
Because there is no cloning step, resequencing to increase accuracy is not possible and it is therefore vital to disentangle noise from true sequence
diversity in this type of data \cite{Quince:2009uq}. Errors can be significantly reduced  when large multiple sequence alignments with close homologs
are available, but in studies of new or not well known organisms, such information is rather sparse. In particular, it is common that there is not enough  similarity to 
differentiate between the sequencing errors and the natural polymorphisms that we want to observe, often leading to artificially inflated diversity estimates~\cite{Kunin2010}.
A few techniques have been developed to remedy to this problem~\cite{Quinlan2008,Medvedev2011} but they do not take
into account all the available information. It is therefore essential to develop methods that can exploit any type of signal available to correct errors.  

In this paper, we introduce \RNApyro, a novel algorithm that enables us to calculate precisely mutational probabilities in RNA sequences with a
conserved consensus secondary structure. We show how our techniques can exploit the structural information embedded in physics-based energy models, 
covariance models and isostericity scales to identify and correct point-wise errors in RNA molecules with conserved secondary structure. In particular, we 
hypothesize that  conserved consensus secondary structures combined with sequence profiles  provide an information that allow us to identify and fix sequencing errors.

Here, we expand the range of algorithmic techniques previously introduced with the \RNAmutants software~\cite{Waldispuhl2008,Waldispuhl2011}.
Instead of exploring the full conformational landscape and sample mutants, we develop an inside-outside algorithm that enables us
to explore the complete mutational landscape with a \emph{fixed} secondary structure and to calculate exactly mutational probability values. In addition
to a gain into the numerical precision, this strategy allows us to drastically reduce the computational complexity ($\mathcal{O}(n^3 \cdot M^2)$ for the
original version of  \RNAmutants to $\mathcal{O}(n \cdot M^2)$ for \RNApyro, where $n$ is the size of the sequence and $M$ the maximal number of mutations).

We design a new scoring scheme combining nearest-neighbor models \cite{Turner2010} to isostericity metrics \cite{Stombaugh2009}.
Classical approaches use a Boltzmann distribution whose weights are estimated using a nearest-neighbour energy model~\cite{Turner2010}. However, the
latter only accounts for  canonical and wobble, base pairs. As was shown by Leontis and Westhof~\cite{Leontis2001},
the diversity of base pairs observed in tertiary structures is much larger, albeit their energetic contribution remains unknown. To quantify geometrical discrepancies, 
an isostericity distance has been designed \cite{Stombaugh2009}, increasing as two base pairs geometrically differ from each other in space. Therefore, we 
incorporate these scores in the Boltzmann weights used by \RNApyro.
 
\TODOTous{Mentionner technologie Illumina 50-75 nts, et modele d'erreur/mutations.}
We illustrate and benchmark our techniques for point-wise error corrections on the 5S ribosomal RNA. We choose the latter since it has been extensively
used for phylogenetic reconstructions~\cite{Hori1987} and its sequence has been recovered for over 712 species (in the Rfam seed alignment with id
\texttt{RF00001}). Using a leave one out strategy, we perform random distributed mutations on a sequence. While our methodology is restricted to the correction of 
point-wise error in structured regions (i.e. with base pairs), we show that \texttt{RNApyro} can successfully extract a signal that can be used to reconstruct the 
original sequence with an excellent accuracy. This suggests that \RNApyro is a promising algorithm to complement existing tools in the NGS error-correction 
pipeline.

The algorithm and the scoring scheme are presented in Sec.~\ref{sec:methods}. Details of the implementation and benchmarks are in Sec.~\ref{sec:results}. 
Finally, we discuss future developments and applications in Sec.~\ref{sec:conclusion}.


\section{Methods}
\label{sec:methods}

We introduce a probabilistic model, which aims at capturing both the stability of the folded RNA and its affinity towards a predefined 3D conformation.
To that purpose, a Boltzmann weighted distribution is assumed, based on a pseudo-energy function $\PE{\cdot}$ which includes contributions for both the free-energy and its putative isostericity towards a multiple sequence alignment. In this model, the probability that the nucleotide at a given position needs to be mutated (i.e. corresponds to a sequencing error) can be computed using a variant of the \emph{Inside-Outside algorithm}~\cite{Lari1990} in time which scales linearly with the sequence length.

\subsubsection{Definitions.}
Let $\B:=\left\{\Ab,\Cb,\Gb,\Ub\right\}$ be the set of nucleotides.
Given an RNA sequence $s\in \B^n$, let $s_i$ be the nucleotide at position $i$. Let $\Omega$ be a set of un-gapped RNA sequences of
length $n$. $\Struct$ is a secondary structure without pseudoknots, denoted by a dot-parenthesis string (well-parenthesized expression with dots). In any such expression, matched parentheses induce and unambiguous set of corresponding positions, associated with base-pairing positions, mediated by hydrogen bonds. It follows that if $(i,j)$ and $(k,l)$ are base pairs in $S$, there is no overlapping extremities  $\{i,j\}\cap \{k,l\}=\varnothing$ and either the intersection is empty 
 ($[i,j]\cap[k,l]=\varnothing$) or one is included in the other ($[k,l]\subset[i,j]$ or 
 $[i,j]\subset[k,l])$. Let us finally denote by $\delta: \B^*\times \B^* \to \mathbb{N}^+$ the Hamming distance, i.e. the number of differing positions between two sequences $s'$ and $s''$ such that $|s'|=|s''|$.
\TODOYann{Move above, define secondary structure and add fancy VARNA illustration}

\subsection{Probabilistic Model}\label{sec:model}
Let $\Omega$ be an gap-free RNA alignment sequence, $S$ its associated secondary structure, 
then any sequence $s$ has probability proportional to its Boltzmann factor
\begin{align*}
  {B}(s) &= e^\frac{-\PE{s}}{RT}, &&\text{with}&\PE{s}&:=\alpha\cdot\ES(s,S)+(1-\alpha)\cdot\EI(s,S,\Omega),
\end{align*}
where $R$ is the Boltzmann constant, $T$ the temperature in Kelvin, $\ES(s)$ and $\EI(s,S,\Omega)$ 
are the free-energy and isostericity contributions respectively (further described below), and $\alpha\in[0,1]$ is an arbitrary parameter that sets the relative weight for both contributions.

\subsubsection{Energy Contribution.}
The free-energy contribution in our pseudo-energy model corresponds to an additive stacking-pairs model, using values from the Turner 2004 model retrieved from the NNDB~\cite{Turner2010}. Given a candidate sequence $s$ for a secondary structure $S$, the free-energy of $S$ on $s$ is given by
\begin{align*}
  \ES(s,S) = \sum_{\substack{(i,j)\to (i',j')\in S\\ \text{stacking pairs}}}\ES^{\beta}_{s_is_j\to s_{i'}s_{j'}} 
\end{align*}
where $\ES^{\beta}_{ab\to a'b'}$ is set to $0$ if $ab=\varnothing$ (no base-pair to stack onto), the tabulated free-energy of stacking pairs $(ab)/(a'b')$ in the Turner model if available, or $\beta\in[0,\infty]$ for non-Watson-Crick/Wobble entries (i.e. neither $\Gb\Ub$, $\Ub\Gb$, $\Cb\Gb$, $\Gb\Cb$, $\Ab\Ub$ nor $\Ub\Ab$). This latter parameter allows to choose whether to simply penalize invalid base pairs, or forbid them altogether ($\beta = \infty$).
The loss of precision due to this simplification of the Turner model remains reasonable since the targeted secondary structure is fixed. For instance, multiloops do not consider base-specific contributions, and therefore their consideration would constitute a criterion for preferring a sequence over another. Furthermore, it greatly eases the design and implementation of dynamic-programming equations. 
\subsubsection{Isostericity Contribution.}
The concept of isostericity score is based on the geometric discrepancy (superimposability) of two base-pairs, using individual additive contributions computed by Stombaugh~\emph{et al}~\cite{Stombaugh2009}. Let $s$ be a candidate sequence for a secondary structure $S$, given in the context of a gap-free RNA alignment $\Omega$,  we define the isostericity contribution to the pseudo-energy as
\begin{align*}
  \ES(s,S,\Omega) &= \sum_{\substack{(i,j)\in S\\ \text{pairs}}}\EI^{\Omega}_{(i,j),s_i s_j}, & \text{where}&& 	\EI^{\Omega}_{(i,j),ab}:=
	\frac{
		\sum_{s'\in\Omega}
			\text{\ISO}((s_i',s_j'),(a,b))}
{		
		|\Omega|
	}
\end{align*}
is the average isostericity of a base-pair in the candidate sequence, compared with the reference alignment.
The $\ISO$ function uses the {Watson-Crick/Watson-Crick} cis isostericity matrix computed by Stombaugh~\emph{et al}~\cite{Stombaugh2009}. Isostericity scores range between $0$ and $9.7$, $0$ corresponding to a perfect isostericity, and a penalty of $10$ is used for missing entries.
The isostericity contribution exponentially favors sequences that are likely to adopt a similar local conformation as the sequences contained in the alignment.
\TODOYann{Avoid duplicate reference to Stombaugh et al}

\subsubsection{Combining contributions.}
Let us remark that any of the individual contributions can be associated to (a subset of) the base-pairs occurring in the structure, possibly complemented, in the case of stacking pairs, with the knowledge of flanking base-pairing nucleotide.
Dropping the implicit dependency on $\Omega$ and $\beta$, let us denote by $\EBP{i,j}{xy}{a'b'}$ the local contribution of a base-pair $(i,j)$ of nucleotides $(a',b')$, surrounded by a stacking pair $(x,y)$ (or $\varnothing$ otherwise), to the pseudo-energy:
\begin{equation}
  \EBP{i,j}{xy}{a'b'}  = \alpha \cdot\ES^\beta_{xy \to a' b'}+(1-\alpha)\cdot\EI^{\Omega}_{(i,j),a'b'}.
\end{equation}

\subsection{Computing the Mutational Profile of Sequences}

Let $s$ be an RNA sequence, $S$ a reference structure, and $m\geq 0$ a desired number of mutations. 
We are interested in  the probability that a given position contains a specific nucleotide, over all sequences having at most $M$ mutations from $s$. Formally, let $\mathcal{D}_{s,M}=\{s'\;|\;\delta(s,s')\le M\}$ be the set of admissible sequences, one aims at computing the probabilities
\begin{equation}
\mathbb{P}(s_i = x\mid s,\Omega, \Struct,M) = \frac{\sum_{\substack{s'\in\mathcal{D}_{s,M}\\\text{s.t. }s'_{i}=x}}B(s')}{\sum_{\substack{s''\in\mathcal{D}_{s,M}}}B(s'')}
\end{equation}

Clearly, the number of sequences in $\mathcal{D}_{s,M}$ grows exponentially with the sequence length, therefore one cannot realistically rely on an exhaustive enumeration to compute the mutational profile. To work around this issue, we propose a linear-time variant of the
 \emph{Inside-Outside algorithm}~\cite{Lari1990} to compute this probability, based on two sets of dynamic programming equations. 

The former, defined in Equations~\eqref{eq:Z_in} and~\eqref{eq:Z_rec_A}-\eqref{eq:Z_rec_C}, is analogous to an \emph{inside} computation: Considering a given substructure of the input structure, it computes the accumulated contributions of any possible sequences that have suitable Hamming distance within the interval. It is therefore similar to a partition function, i.e. the sum of Boltzmann factors over all sequences within $[i,j]$, 
knowing that position $i-1$ is composed of nucleotide $a$ (resp. $j+1$ is $b$), within 
$m$ mutations of $s$. 

The latter, defined by Equations~\eqref{eq:Y_in} and~\eqref{eq:Y_rec_A}-\eqref{eq:Y_rec_D},
 computes the \emph{outside} algorithm,   
the partition function over sequences within $m$ mutations of $s$ outside the interval (restricted to two intervals $[0,i]\cup[j,n-1]$), 
knowing  that flanking inner positions $(i+1,j-1)$ contain nucleotides $a$ and $b$ respectively. A suitable combination of these terms, computed as shown in Equations~\eqref{eq:combine_A}-\eqref{eq:combine_C}, gives the total weight of every possible sequences that support a given base-pair (or unpaired position) and, in turn, the probability of seeing a specific base at a given position.


\subsubsection{Inside computation.}
\begin{figure}[t]\centering
\includegraphics[scale=\ScaleDP]{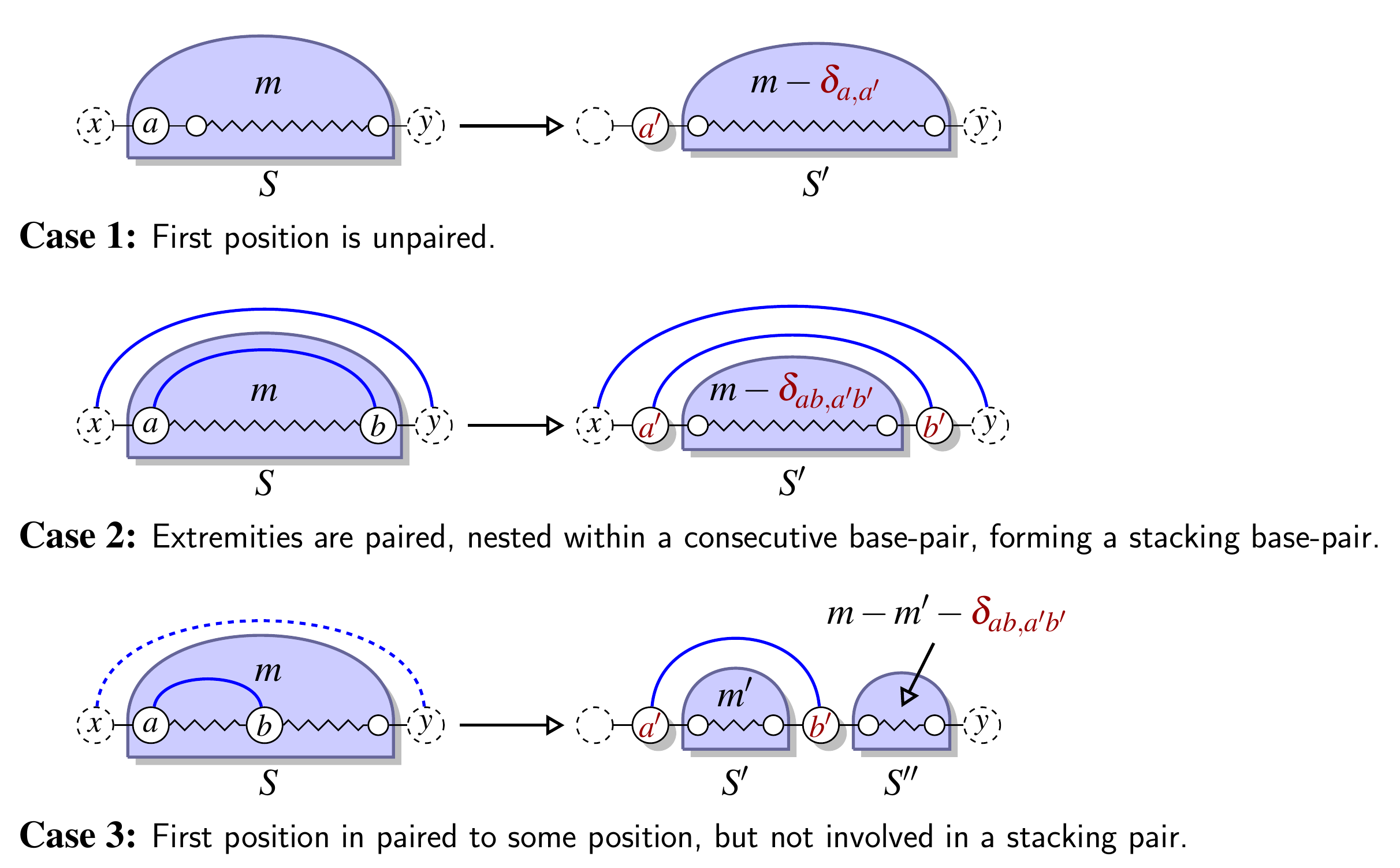}
\caption{Principle of the inside computation (partition function). Any sequence (mutated)  
can be decomposed as a sequence preceded by a, possibly mutated, base 
(Unpaired case), a sequence surrounded by some base-pair (Stacking-pair case), 
or as two sequences segregated by some base-pair (General base-pairing case). In this latter case, mutations must be distributed between sub-sequences.\label{fig:inside}}
\end{figure}

The \emph{Inside} function $\Z{\Struct}{m}{x,y}$ is simply the partition function, i.e. the sum of Boltzmann factors over all sequences for a substructure $\Struct$ (implicitly attached to an interval $[i,j]$ of the sequence), featuring $m$ mutations/errors compared to $s$, and having flanking nucleotides $x$ and $y$. 
Such terms can be computed recursively, using the following equation for the initial case:
\begin{equation}
	\forall x,y\in \B\times \B, m\in [0,M]:\, \Z{\varepsilon}{m}{x,y}=\left\{
	\begin{array}{ll}
		1 &\text{ If } m = 0\\
		0 &\text{ Otherwise.}
	\end{array}\right.
\label{eq:Z_in}
\end{equation}
In other words, either there is no sequence at distance $m>0$ of the empty sequence, or the only allowed sequence is the empty sequence ($m=0$), having energy $0$. Since the contributions only depend on base pairs, they do not appear in the initial conditions. 

The main recursion considers a general structure $\Struct$, flanked by two nucleotides (outside the region of interest) $x$ and $y$, respectively on the $5'$ and $3'$ end of the sequence. As illustrated by Figure~\ref{fig:inside}, it is computed, for each subinterval $[i,j]$, by considering one of the three following cases, dependent on the base-pairing status and context of the leftmost position in the sequence/structure:
\begin{itemize}
\item {\bf Case 1: Unpaired leftmost position.} If the first position  is unpaired in the structure, then $\Struct$ can be further decomposed, using a dot-parenthesis notation, as $\Struct = \ub \Struct'$. Let $a\in B$ be the nucleotide found at the leftmost position in the initial structure, then one has:
\begin{equation}
\label{eq:Z_rec_A}
	\Z{\Struct}{m}{x,y} =
      \sum_{\substack{a'\in \B,\\ \Kron_{a,a'}\le m}}  
      \Z{\Struct'}{m-\Kron_{a,a'}}{a',y}.
\end{equation}
Indeed, any suitable sequence is a concatenation of a, possibly mutated, nucleotide $a'$ at the first position, followed by a sequence over the remaining interval, having $m-\Kron_{a,a'}$ mutations (accounting for a possible mutation at first position), and having flanking nucleotides $a'$ and $y$.
\item {\bf Case 2: Paired ends, stacking onto another base-pair.} If both ends of the considered interval form a base-pair ($\Struct = \op\Struct'\cp$), stacking onto another base pair just outside whole region, then the isosteric contribution of the base-pair must be supplemented with a specific "stacking-pairs" bonus. Let $a$ and $b$ be the nucleotides found on both ends of the interval (positions $i$ and $j$), then one has
\begin{equation}
\label{eq:Z_rec_B}
	\Z{\Struct}{m}{x,y} =
      \sum_{\substack{a',b'\in \B^2,\\ \Kron_{ab,a'b'}\le m}}
			 e^{\frac{-\EBP{i,j}{xy}{a'b'}}{RT}}
			 \cdot \Z{S'}{m-\Kron_{ab,a'b'}}{a',b'}.
\end{equation}
Any sequence generated here consists of two, possibly mutated, nucleotides $a'$ and $b'$, flanking a sequence over the remaining portion. In order for the total distance to sum to $m$, this portion must feature $m-\Kron_{ab,a'b'}$ additional point wise mutations.

\item {\bf Case 3: Paired leftmost position, but no stacking pairs.} In this case, the structure is split into two parts by the base pair ($\Struct = \op \Struct'\cp\Struct''$). Let us denote by $k$ the partner of position $i$, and by $a$, $b$ and $c$ the bases found at positions $i$, $k$ and $j$ respectively, then one has:
\begin{equation}
\label{eq:Z_rec_C}
	\Z{\Struct}{m}{x,y}=\sum_{\substack{a',b'\in \B^2,\\ \Kron_{ab, a'b'}\le m}}
      \sum_{m'=0}^{m-\Kron_{ab,a'b'}}
   		 e^{\frac{-\EBP{i,k}{\varnothing}{a'b'}}{RT}}
      \cdot\Z{\Struct'}{m-m'-\Kron_{ab,a'b'}}{a',b'}
      \cdot\Z{\Struct''}{m'}{b',y}.
\end{equation}
In other words, if the leftmost position is paired, and the base-pair is not stacking onto another base-pair, then the  only term contributing directly to the energy is the isostericity of the base pair. Admissible sequences for $\Struct$ consist of two paired nucleotides $a'$ and $b'$ at positions $i$ and $k$ respectively, flanking a sequence for $\Struct'$ (over an interval $[i+1,k-1]$), and followed by a (possibly empty) sequence for $\Struct''$ (over $[k+1,j]$). Since the total number of mutations sums to $m$, a parameter $m'$ is introduced to distribute the remaining mutations between the two sequences.
\end{itemize}

\subsubsection{Outside computation.}	

The \emph{Outside} function, $\Y{\Struct}{m}{x,y}$ is the partition function considering only the 
contributions of subsequences excluding a given structure/interval $\Struct$, occupying the open interval $]i,j[$ in the sequence, at Hamming distance exactly $m$ to the initial sequence $s$, and assuming that nucleotides $x$ and $y$ were previously chosen for $i+1$ and $j-1$, the outermost portions of the excluded structure.
The associated terms $\Y{\Struct}{m}{x,y}$ can then be computed recursively, initially considering the case of any prefix $\Struct'$ of the complete structure:
\begin{equation}
	\forall x,y\in \B\times \B, \forall \Struct'\text{ s.t. }\Struct=\Struct'.\Struct'', m\in [0,M]:\, \Y{\Struct'}{m}{x,y}=\Z{\Struct''}{m}{y,z}
\label{eq:Y_in}
\end{equation}
where $z\in\B$ can be any nucleotide, and provably does not affect further computations. 
In other words, the sequences explored by an outside computation, excluding a prefix of $\Struct$, are exactly the sequences generated on the corresponded suffix. This set of sequence is also the inside term on the suffix structure.
It is also worth pointing out that $\Y{\Struct}{m}{x,y}= \Z{\varepsilon}{m}{y,X}=\mathbb{1}_{m=0}$.

\begin{figure}[t]\centering
\includegraphics[scale=\ScaleDP]{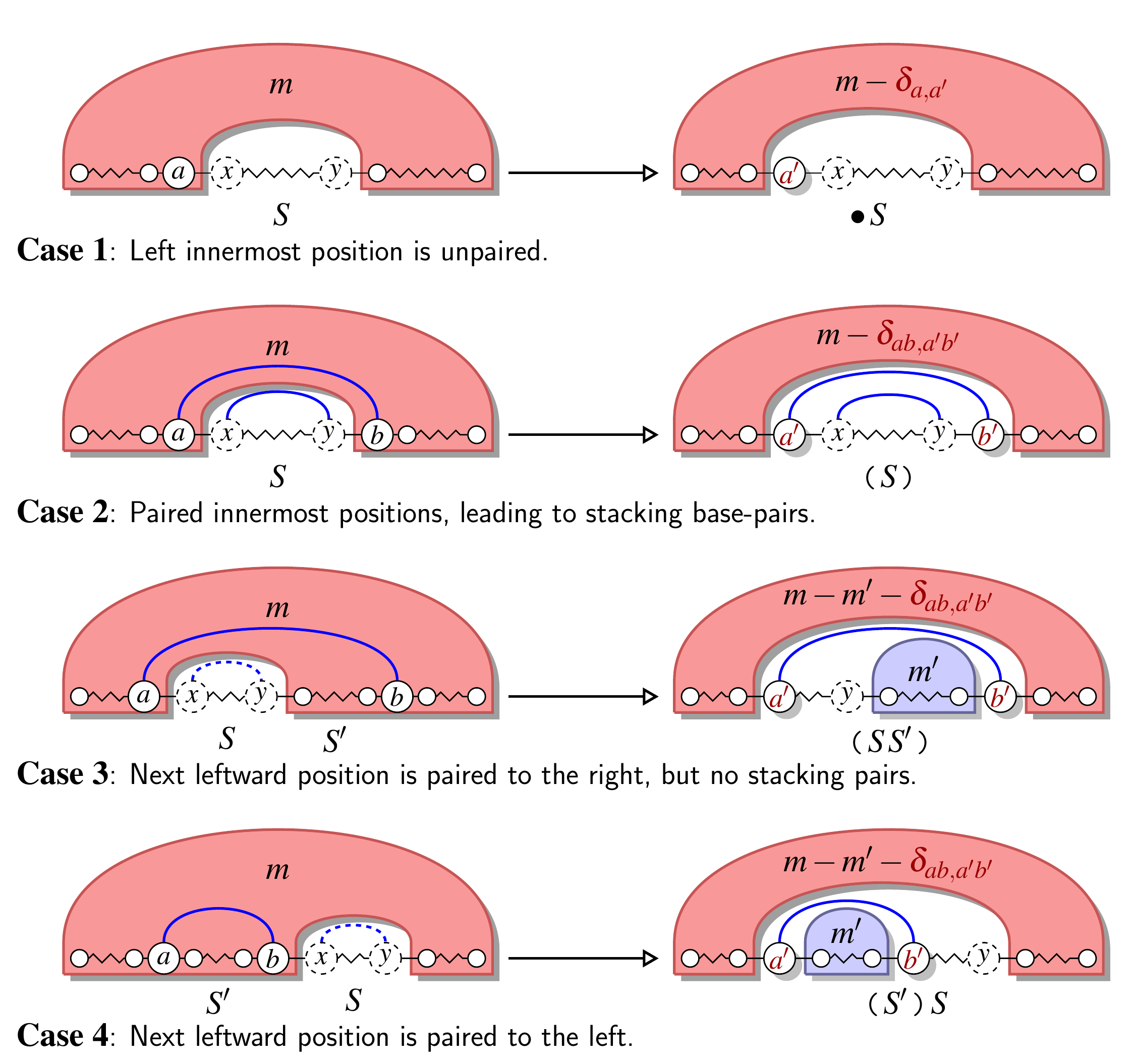}
\caption{Principle of the outside computation. Note that the outside algorithm uses intermediate results from the inside algorithm, 
therefore its efficient implementation requires the precomputation and storage of the inside contributions.\label{fig:outside}}
\end{figure}
In general, the main recurrence below works by extending the excluded structure $\Struct$ (covering the $]i,j[$ interval) in the leftward direction. As shown in Figure~\ref{fig:outside}, four cases must be considered, depending on the base-pairing status and directionality of the next considered position:
\begin{itemize}
\item {\bf Case 1: Unpaired position.} When the innermost leftward base is unpaired, any base $a\in\B$ may be chosen for this position, and the rest of the sequence is generated recursively. The number of remaining mutations may be decreased by the choice of $a$, leading to the following recurrence:
\begin{equation}
	\Y{\Struct}{m}{x,y} = \sum_{\substack{a'\in \B,\\ \Kron_{a,a'}\le m}}
    \Y{\ub\Struct}{m- \Kron_{a,a'}}{a',y}
\label{eq:Y_rec_A}
\end{equation}
\item {\bf Case 2: Stacking base-pair.} If both ends of the excluded structure are paired together, and are nested within another base-pair in the remaining structure, then an additional contribution, stemming from a stacking pairs energy, has to be considered. The outside terms are then computed by simulating any pair of base-pairing nucleotides for $i,j$, and by proceeding recursively on the remaining portion, as follows:
\begin{equation}
	\Y{\Struct}{m}{x,y} = 
    \sum_{\substack{a'b'\in \B^2,\\ \Kron_{ab,a'b'}\le m}}
		 e^{\frac{-\EBP{i,j}{xy}{a'b'}}{RT}}\cdot
    \Y{\op\Struct\cp}{m- \Kron_{ab,a'b'}}{a',b'} 
\label{eq:Y_rec_B}
\end{equation}
\item {\bf Case 3: Next position paired rightward, in the absence of stacking pair.} In this case, the innermost leftward position $i$ is paired to the right at some position $k$ . 
Let us assume that its partner resides outside the excluded structure $\Struct$ (This assumption is provably without loss of generality, and directly follows from the fact that $\Struct$ is well-parenthesized). 
The structure within the base-pair may be described by the expression $\op\Struct \Struct'\cp$, where $\Struct'$ may possibly be empty (except if $\Struct$ is enclosed within corresponding brackets, to avoid stacking pairs). Therefore, any sequence considered by the outside computation consists in three independent parts: two nucleotides for the paired positions, a sequence for the region excluding $\Struct''$, and a sequence for $\Struct'$. It follows that:
\begin{equation}
	\Y{\Struct}{m}{x,y} = \sum_{\substack{a'b'\in \B^2,\\ \Kron_{ab,a'b'}\le m}}
		 \sum_{m'=0}^{m-\Kron_{ab,a'b'}}
  		 e^{\frac{-\EBP{i,k}{\varnothing}{a'b'}}{RT}}
		 \cdot\Y{\op\Struct \Struct'\cp}{m- \Kron_{ab,a'b'} - m'}{a',b'}
     \cdot\Z{\Struct'}{m'}{y,b'} .
\label{eq:Y_rec_C}
\end{equation}
\item {\bf Case 4: Next position paired leftward.} If the innermost, leftward, position $i$ is paired to some position $k<i$, delimiting a substructure $\Struct'$, then any sequence considered by the outside computation consists in three parts: two nucleotides $a$ and $b$, a sequence for $\Struct'$, and a sequence for the region excluding $\op\Struct'\cp\Struct$. Consequently, one has:
\begin{equation}
	\Y{\Struct}{m}{x,y} = 
		 \sum_{\substack{a'b'\in \B^2,\\ \Kron_{ab,a'b'}\le m}}
		 \sum_{m'=0}^{m-\Kron_{ab,a'b'}}
   	 e^{\frac{-\EBP{k,i}{\varnothing}{a'b'}}{RT}}
		 \cdot\Y{\op\Struct'\cp\Struct}{m- \Kron_{ab,a'b'} - m'}{a',b}
     \cdot\Z{\Struct'}{m'}{a',b'}
\label{eq:Y_rec_D}
\end{equation}
\end{itemize}

\subsubsection{Combining Inside and Outside Computations into Point-Wise Mutations Probabilities.}

We are now left to compute the probability that a a given nucleotide $a\in\B$ is found at a given position $i$.
This quantity can also be expressed as the ratio of $\mathcal{W}^M_{\substack{i, [a]}}$, the total Boltzmann weight 
of the set of sequences featuring the nucleotide, and $\Z{\varepsilon}{\le M}{X,X}$ the total weight of sequences having at most $M$ mutations:
\begin{equation}
	\mathbb{P}(s_i = a\;|\; M) := \frac{\mathcal{W}^M_{\substack{i, [a]}}}{\Z{\varepsilon}{\le M}{X,X'}} = \frac{\mathcal{W}^M_{\substack{i, [a]}}}{\sum_{m=0}^{M}\Z{\varepsilon}{m}{X,X'}}\label{eq:normalize}
\end{equation}
where $X,X'\in\B$ may be any nucleotides (no impact on energy/weights).

\begin{figure}[t]\centering
\includegraphics[scale=\ScaleDP]{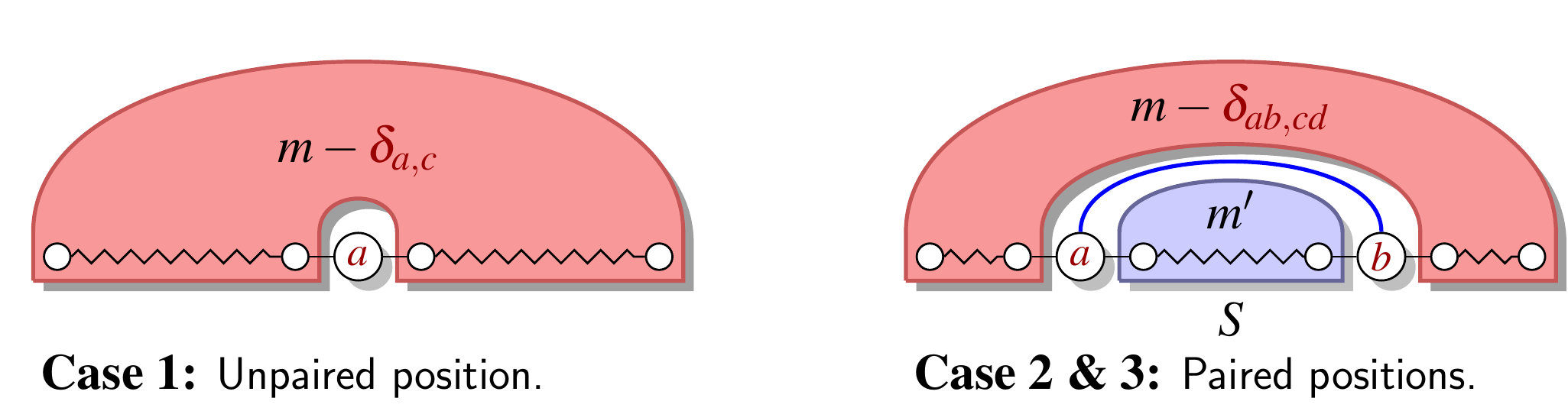}
\caption{Combining the inside and outside contributions to compute the total Boltzmann weight of all sequences having a given base (case 1) or base-pair (cases 2 and 3) at a given position.\label{fig:combine}}
\end{figure}

To that purpose, one leverages the \emph{Inside-Outside} construction, as  illustrated by Figure~\ref{fig:combine}. Namely, while computing the total contribution of sequences featuring a base $a'\in\B$ at position $i$, one must consider the three following situations:
\begin{itemize}
\item {\bf Case 1: Unpaired position.} In this case, the supporting sequences are simply those excluding the structure $\ub$ at position $i$, where a base $c\in\B$ was formerly found. Summing over any possible number of mutations for the \emph{outside} region, one obtains:
\begin{equation}
 \mathcal{W}^M_{\substack{i, [a]}} =  \sum_{m=\Kron_{a,c}}^{M}
			\Y{\ub}{m-\Kron_{a,c}}{a,a}.
\label{eq:combine_A}
\end{equation}
\item {\bf Case 2: Position is paired rightward.} Here, position $i$ is paired to some position $k>i$, whose content impacts the base-pair contribution. Any sequence having $a$ at position $i$ can be decomposed as a base-pair, nesting a sequence for the structure $\Struct'$ on the interval $[i+1,k-1]$, and an outside sequence, excluding the structure $\op\Struct'\cp$.
Consequently, let $c$ and $d$ be the original nucleotides at positions $i$ and $k$, one has:
\begin{equation}
 \mathcal{W}^M_{\substack{i, [a]}} =  
			\sum_{m=0}^{M}
			\sum_{\substack{b\in \B\\\Kron_{ab,cd}\leq m}}
			\sum_{m'=0}^{m-\Kron_{ab,cd}}
     	 e^{\frac{-\EBP{i,k}{\varnothing}{ab}}{RT}}
			\cdot\Y{\op\Struct'\cp}{m-\Kron_{ab,cd}}{a,b}
			\cdot\Z{\Struct'}{m'}{a,b}
\label{eq:combine_B}
\end{equation}
\item {\bf Case 3: Position is paired leftward.} This case is symmetrical to the previous one, with the exception that $k<i$.
Consequently, let $c$ and $d$ be the original nucleotides at positions $i$ and $k$, one has:
\begin{equation}
 \mathcal{W}^M_{\substack{i, [a]}} =  
			\sum_{m=0}^{M}
			\sum_{\substack{b\in \B\\\Kron_{ba,cd}\leq m}}
			\sum_{m'=0}^{m-\Kron_{ba,cd}}
     	 e^{\frac{-\EBP{k,i}{\varnothing}{ba}}{RT}}
			\cdot\Y{\op\Struct'\cp}{m-\Kron_{ba,cd}}{b,a}
			\cdot\Z{\Struct'}{m'}{b,a}
\label{eq:combine_C}
\end{equation}
\end{itemize}

\subsection{Complexity Considerations}
Using dynamic programming, Equations~\eqref{eq:Z_rec_A}-\eqref{eq:Z_rec_C} and~\eqref{eq:Y_rec_A}-\eqref{eq:Y_rec_D} can be computed in linear time and memory. Namely, the $\mathcal{Z}^{*}_{*}$ and $\mathcal{Y}^{*}_{*}$ terms are computed starting from smaller values of $m$ and structure lengths, storing the results as they become available to ensure constant-time access during later stages of the computation. Furthermore, energy terms $E(\cdot)$ may be accessed in constant time after a simple precomputation of the isostericity contributions in $\Theta(n\cdot|\Omega|)$. Computing any given term therefore requires $\Theta(m)$ operations due to the explicit distribution of the number of mutations.

In principle, $\Theta(M\cdot n^2)$ terms, identified by different $(\Struct,m)$ triplets, should be computed.
However, a close inspection of the recurrences reveals that the computation can be safely restricted to a subset of intervals $(i,j)$.
For instance, the inside algorithm only requires computing intervals $[i,j]$ that do not break any base-pair, and whose next position $j+1$ is either past the end of the sequence, or is base-paired prior to $i$. A similar property holds for the outside computation,  following from the linearity of the outside recurrence (i.e. the computation of the outside term for a given excluded structure only relies on the computation of another, strictly larger, structure).
 
These properties drastically limit the combinatorics of required computations, dropping from $\Theta(n^2)$ to $\Theta(n)$ the number of terms that need to be computed and stored. Consequently the overall complexity of the algorithm is $\Theta(n\cdot(|\Omega|+M^2))$ arithmetic operations and $\Theta(n\cdot(|\Omega|+M))$ memory.

\section{Results}
\label{sec:results}

\subsection{Implementation}
\begin{figure}[t]
{\centering \includegraphics[width=.5\linewidth]{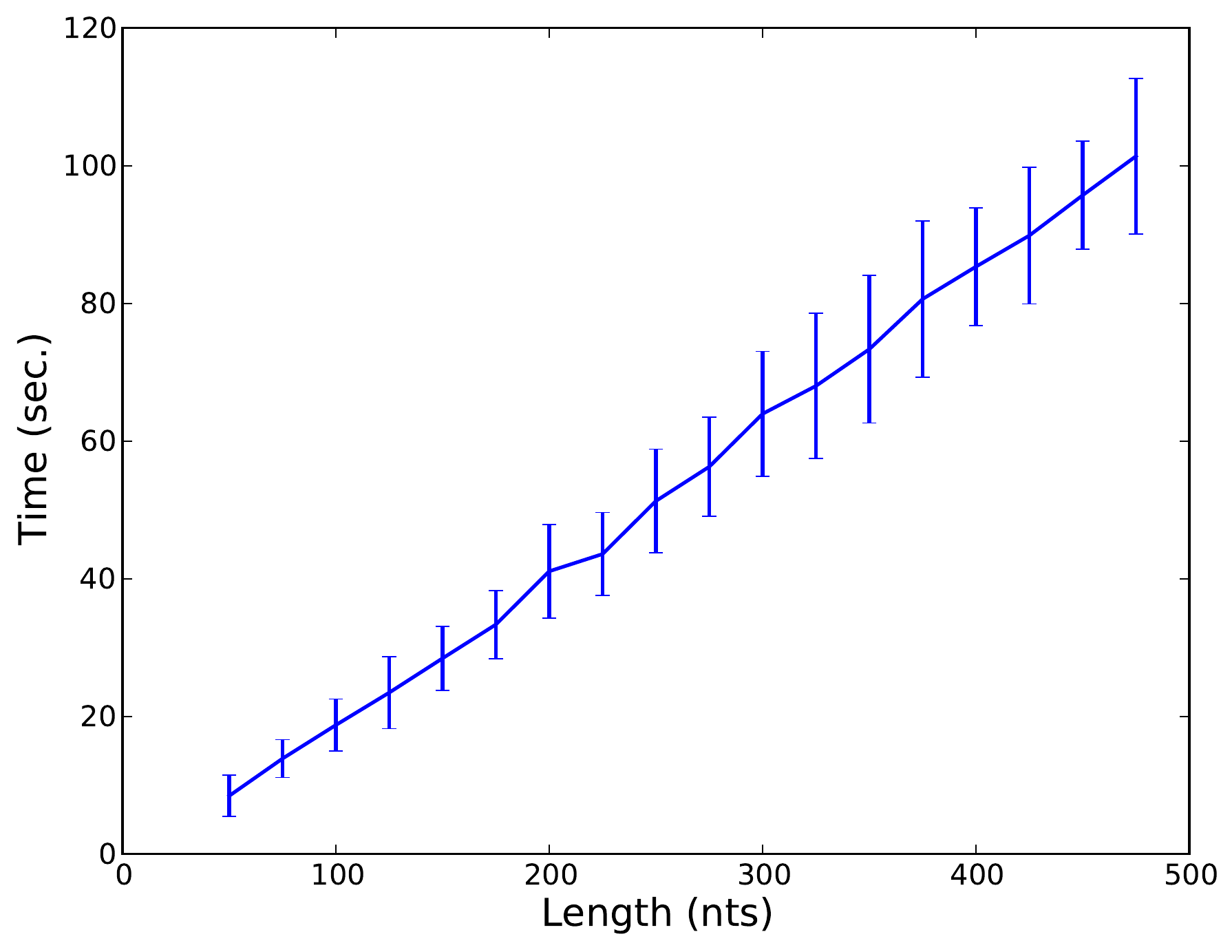}\\}

\caption{Typical runtimes required by the computation of mutational profiles 
averaged on $50$ random sequences for each length ranging from $100$ and $400$ nts, while allowing for a maximal number mutations equal to $M=10$. 
For each sequence, a random multiple sequence alignment was generated, consisting of $51$ aligned sequences, each compatible with a randomly generated consensus secondary structure. 
}
\label{fig:time}
\end{figure}

The software was implemented in Python2.7 using the \textit{mpmath}~\cite{mpmath} library
for  arbitrary floating point precision. The source code is freely available at:


{\centering \url{http://csb.cs.mcgill.ca/RNApyro}\\}
The time benchmarks were performed on an AMD Opteron(tm) Processor 6278  at 2.4 GHz with cache of 512 KB.20 cores and 175 GB ram.
Since typical use-cases of \RNApyro require efficiency and scalability, we present in TableFigure~\ref{fig:time}
typical runtimes required to compute the probabilities for  every nucleotide at every positions for a vast set of parameters.
For those tests, both the multiple sequence alignment and the reference sequence were generated uniformly at random,
based on a realistic random secondary structure, generated as described in Levin~\emph{et al}~\cite{Levin:2012kx}.
\TODOVlad{Add more extensive time benchmark, as a plot}

\subsection{Homogenous Error Correction in 5s rRNAs}
\label{sec:5S}
To illustrate the potential of our algorithm, we applied our techniques to identify and correct point-wise errors in RNA sequences
with conserved secondary structures. More precisely, we used \RNApyro to reconstruct 5s rRNA sequences with randomly distributed
mutations. This experiment has been designed to suggest further applications to error-corrections in pyrosequencing data.

We built our data set from the 5S rRNA multiple sequence alignment (MSA) available in the Rfam Database 11.0 (Rfam id: \texttt{RF00001}).
Since our software does not currently implement gaps (mainly because scoring indels is a challenging issue that cannot be fully addressed
in this work),  we clustered together the sequences with identical gap locations. From the $54$ MSAs without gap produced, we selected the
biggest MSA  which contains $130$ sequences (out of $712$ in the original Rfam MSA). Then, in order to avoid overfitting, we used \texttt{cd-hit}
\cite{Li:2006fk} to remove sequences with more than 80\% of sequence similarity. This operation resulted in a data set of $45$ sequences. 

We designed our benchmark using a leave-one-out strategy. We randomly picked a single sequence from our data set and performed $12$ random
mutations, corresponding to an error-rate of 10\%. We repeated this operation $10$ times. The value of $\beta$ was set to $15$ (larger values gave similar results). 
To estimate the impact on the distribution of the relative weights of energy and isostericity, we used 4 different values of $\alpha = {0, 0.5, 0.8, 1.0}$. 
Similarly, we also investigated the impact of an under- and over- estimate of the number of errors, by setting the presumed number of errors to 50\% (6 mutations) and 200\% (24 mutations) of their exact number (i.e. $12$).

\begin{figure}
\centering
	\includegraphics[width=\textwidth]{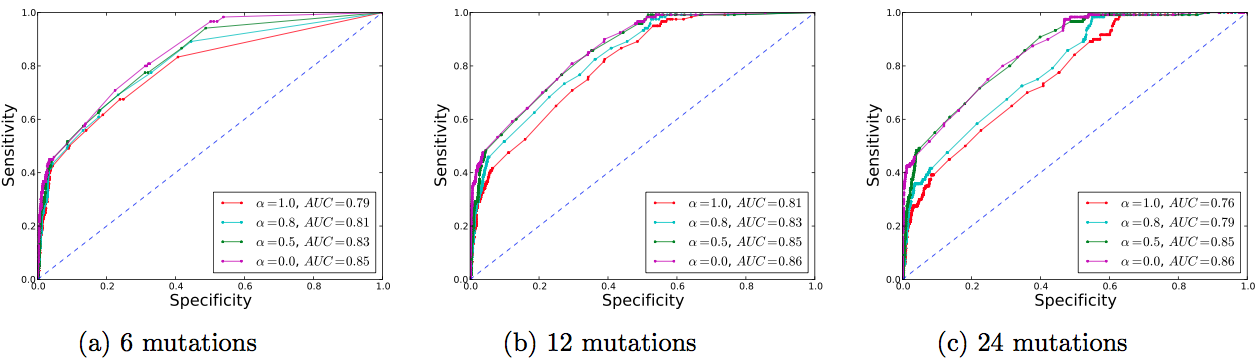}\\

\caption{Performance of error-correction. Subfigures show accuracy with under-estimated error rates (6 mutations), exact estimates (12 mutations) and over estimates 
(24 mutations). We also analyze the impact of the parameter $\alpha$ distributing the weights of stacking pair energies vs isostericity scores and use values 
ranging of $\alpha=\{0,0.5,0.8,1.0\}$. The AUC is indicated in the legend of the figures. Each individual ROC curve represent the average performance over the 10 experiments.}
\label{fig:ROCall}\SpaceCheating
\end{figure}

To evaluate our method, we computed a ROC curve representing the performance of a classifier based on the mutational probabilities computed
by \RNApyro. More specifically, we fixed a threshold $\lambda \in [0,1]$, and predicted an error at position $i$ in sequence $\omega$ if and only if the
probability $P(i,x)$ of a nucleotide $x$ exceeds this threshold. To correct the errors we used the set of nucleotides having probability
greater than $\lambda$, that is  
$$C(i) = \{ x \; | \;  x \in \{ \Ab,\Cb,\Gb,\Ub \}, P(i,x) > \lambda \mbox{ and }  n \neq \omega[i] \},$$
 where $\omega[i]$ is
the nucleotide at position $i$ in the input sequence. We note that, for lower thresholds, multiple nucleotides may be available in $C(i)$ to correct
the sequence. Here, we remind that our aim is to estimate the potential of error-correction of \RNApyro, and not to develop a full-fledged error-correction pipe-line, which  
shall be the subject of further studies. Finally, we progressively varied $\lambda$ between $0$ and $1$ to calculate the ROC curve and the area
under the curve (AUC). Our results are reported in Figure~\ref{fig:ROCall}. 

Our data demonstrates that our algorithm shows interesting potential for error-correction applications. First, the AUC values (up to $0.86$) indicate that a
signal has been successfully extracted. This result has been achieved with errors in loop regions (i.e. without base pairing information) and thus suggests
that correction rates in structured regions (i.e. base paired regions) could be even higher. Next, the optimal values of $\alpha$ tend to be close to $0.0$. This 
finding suggests that, at this point, the information issued from the consideration of stacking energies is currently modest. However, specific examples showed improved performance
using this energy term. Further studies must be conducted to understand how to make the best use of it. Finally, our algorithm seems robust to the number of
presumed mutations. Indeed, good AUC values are achieved even with conservative estimates for the number of errors (c.f.~50\% of the errors, leading to 
Fig.~\ref{fig:ROCall}(a)), as well as with large  values (cf~200\% of the errors  in Fig.~\ref{fig:ROCall}(c)). It is worth noting that scoring schemes giving a larger weight on
the isostericity scores (i.e. for low $\alpha$ values) seem more robust to under- and over-estimating the number of errors.

\subsection{Correcting Illumina sequencing errors in 16s rRNAs}
\label{sec:16S}
\TODOTous{Describe and present results????}

To complete our benchmark, we turned to the small subunit ribosomal RNA in bacteriae, a molecule which is of particular interest  in metagenomics and phylogenetics. Our aim was to get as close as possible to a pyrosequencing context, in which reads are produced non-uniformly by an Illumina sequencer, impacting the distribution of errors in the sequence. 
We chose this setting both because of the popularity of Illumina sequencing in metagenomics, and since the underlying sequencing technique only considers base substitutions (no insertions), the only type of errors currently detected by \RNApyro.
To that purpose, we used simulated Illumina reads, mapped the reads back to a reference alignment, and run \RNApyro on a consensus sequence derived from the mapped reads, estimating the maximal amount of mutations from both the length of the sequence and the sequencing depth.

We gathered the seed sequences of the bacterial multiple sequence alignment (MSA) retrieved from the RFAM Database 11.0 (Rfam id: RF00177)~\cite{gardner2011rfam}. This alignment is composed of 93 sequences, whose length varies between 1461 and 1568 nucleotides, and has an average pairwise sequence identity of $69\%$. We used the pseudoknot-free version of the consensus secondary structure. A secondary structure for a specific reference sequence was obtained by simply mapping the structure back to sequence, i.e. by removing any base-pair having at least one partner involved in a gapped position. For similar reasons, we locally excluded from our calculation of the isostericity contribution for a given base-pair, described in Section~\ref{sec:model}, any sequence featuring at least a gap on the corresponding positions.

To simulate sequencing errors, we used the next-generation sequencing read simulator 
{\tt ART}~\cite{huang2012art}. The \emph{Illumina technology} setting was chosen as the main error mode, generating reads of $75~\text{bps}$, featuring mostly base substitution errors. Reads were mapped back to the original sequence, and a consensus sequence was determined, from the sequencing output, by a simple majority vote in the case of multiple coverage. Uncovered regions were simply generated at random. The average  rate of errors observed in the final consensus sequence was empirically estimated to represent $2.4\%$, $0.9\%$ and $0.6\%$ of the reference sequence, for prescribed sequencing coverages of $5$, $10$ and $15$ fold respectively.

\begin{figure}
\centering
\begin{subfigure}{.33\textwidth}
  \centering
  \includegraphics[width=\linewidth]{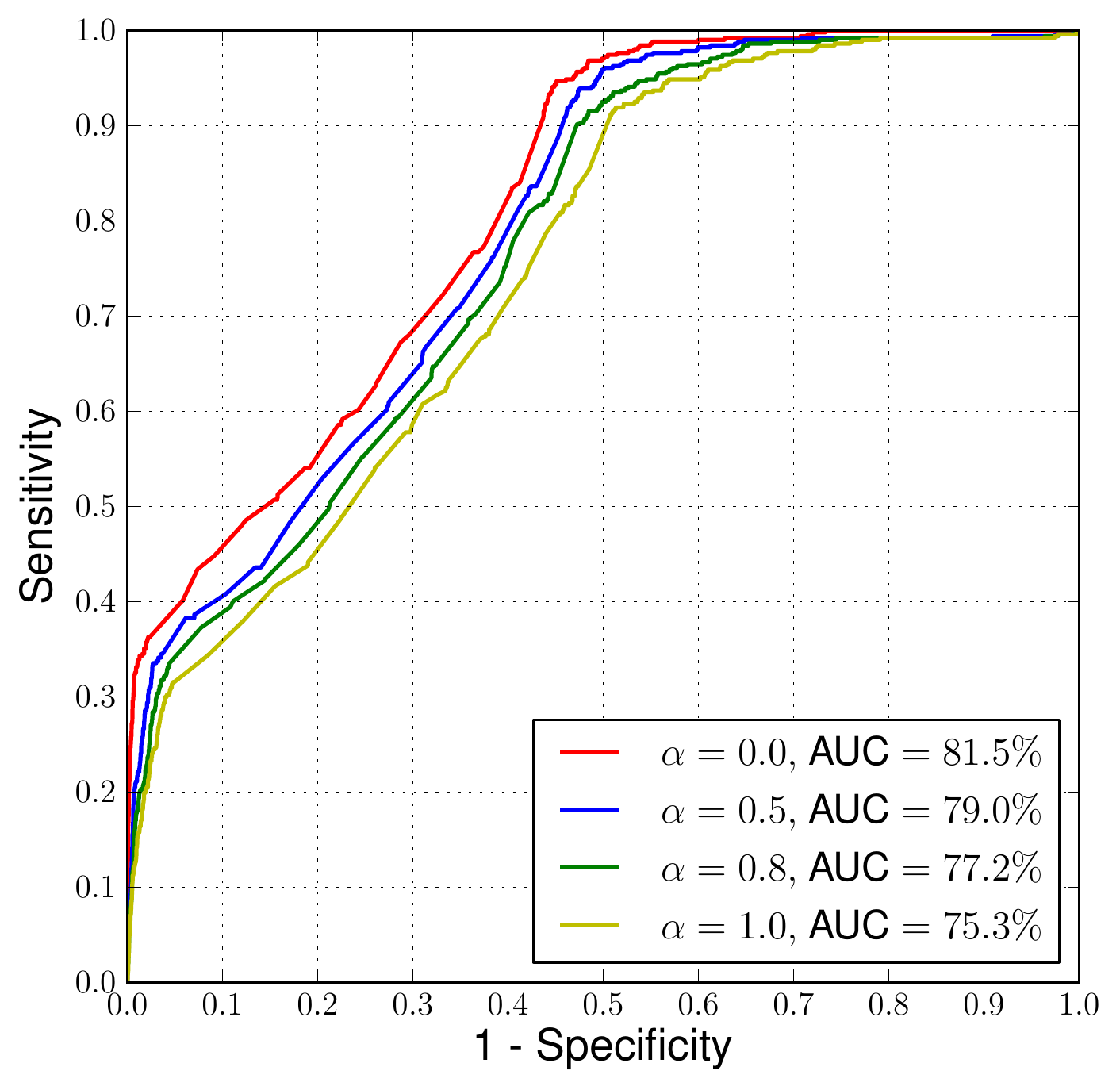}
  \caption{$5-$fold coverage}
\end{subfigure}%
\begin{subfigure}{.33\textwidth}
  \centering
  \includegraphics[width=\linewidth]{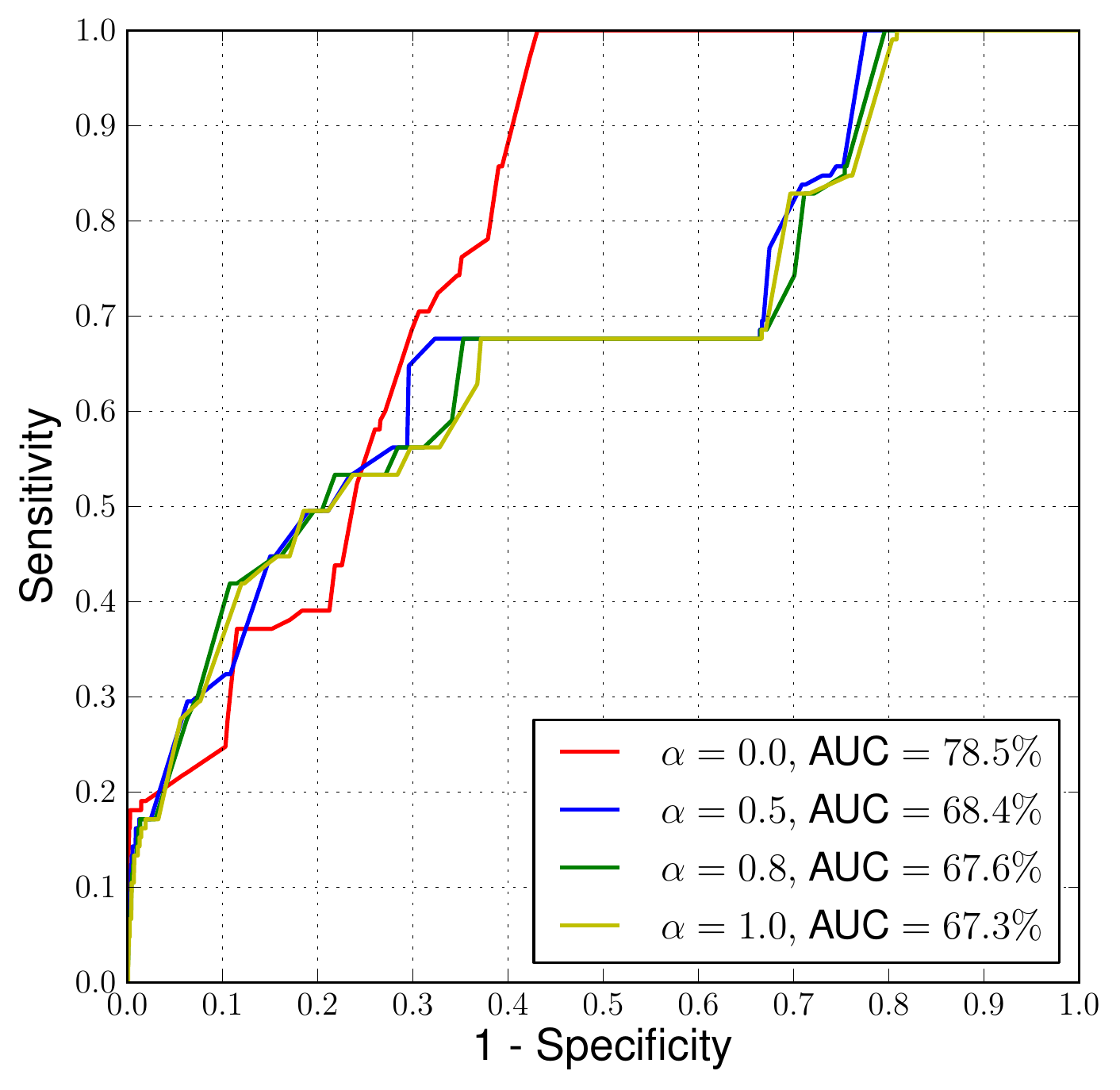}
    \caption{$10-$fold coverage}
\end{subfigure}
\begin{subfigure}{.33\textwidth}
  \centering
  \includegraphics[width=\linewidth]{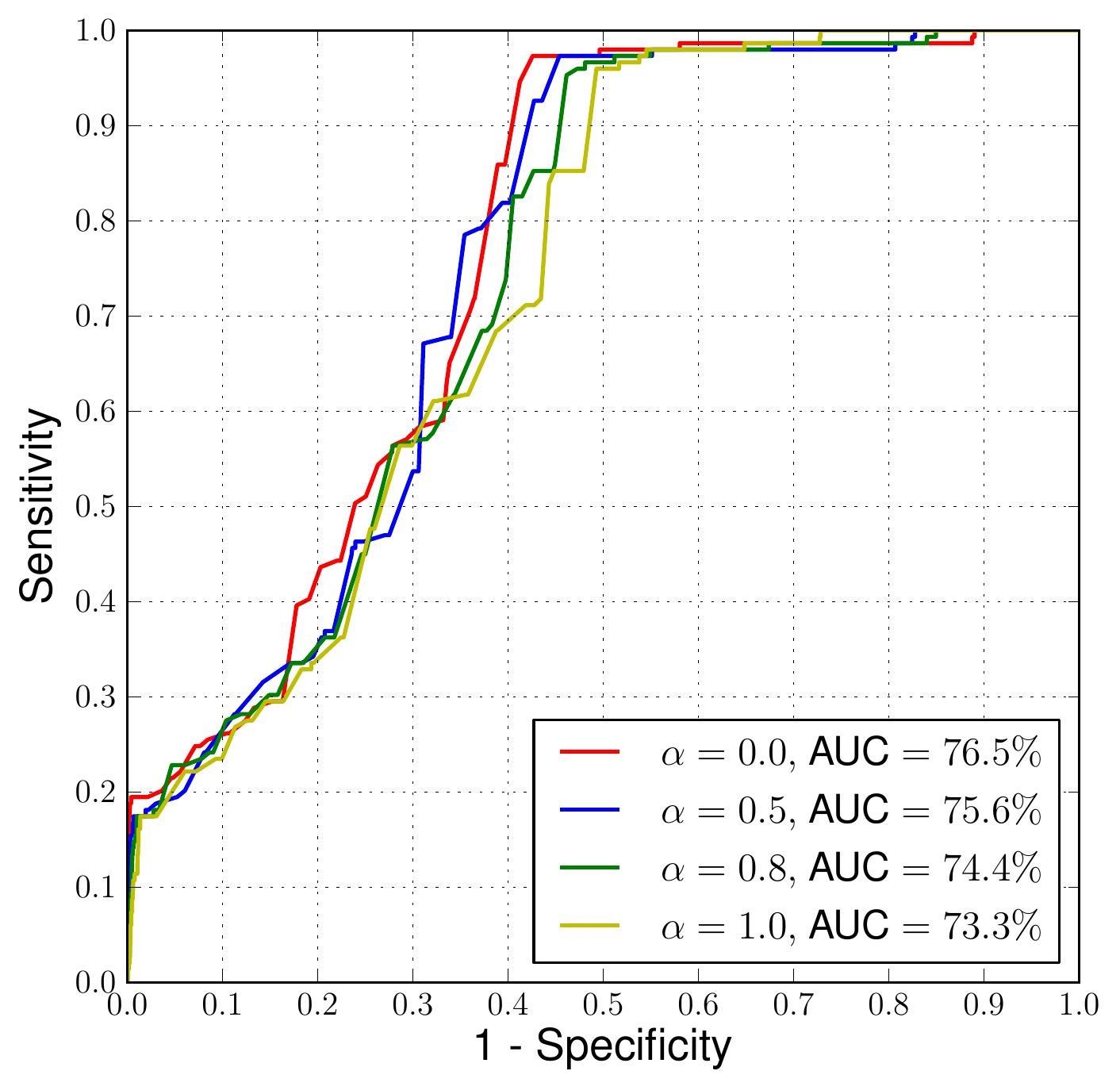}
    \caption{$15-$fold coverage}
\end{subfigure}
\caption{Performance of error-correction over all positions. Subfigures show accuracy when
ART fold parameter is set to 5, 10 and $15-$fold coverage. We also analyze the impact of the parameter $\alpha$ distributing the weights of stacking pair energies vs isostericity scores and use values ranging of $\alpha = \{0, 0.5, 0.8, 1.0\}$. The AUC is indicated in the legend of the figures. Each individual ROC curve represent the average performance over at least 10 experiments.}

\label{fig:16s_all}
\end{figure}

\begin{figure}
\centering
\begin{subfigure}{.33\textwidth}
  \centering
  \includegraphics[width=\linewidth]{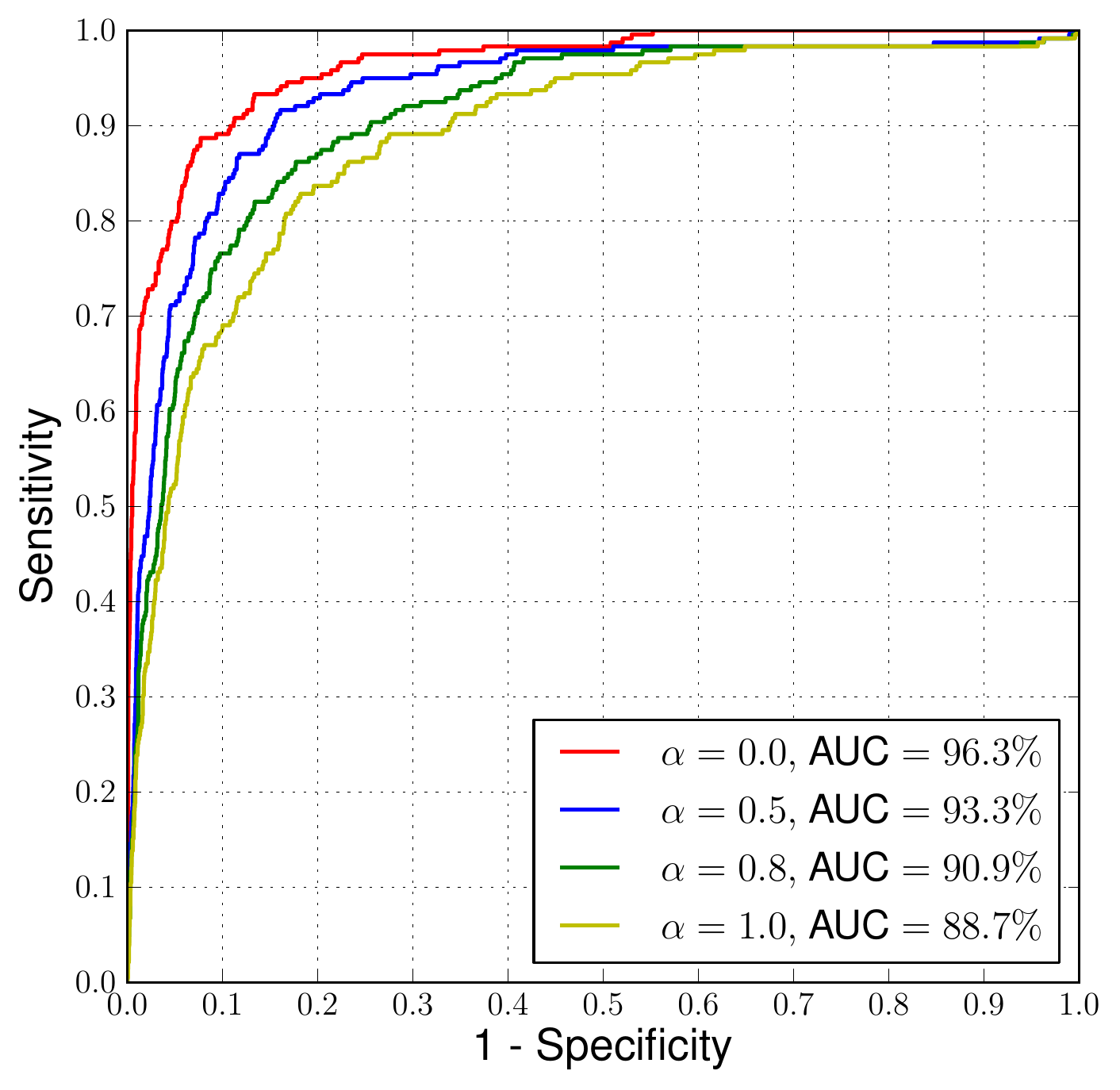}
  \caption{$5-$fold coverage}
\end{subfigure}%
\begin{subfigure}{.33\textwidth}
  \centering
  \includegraphics[width=\linewidth]{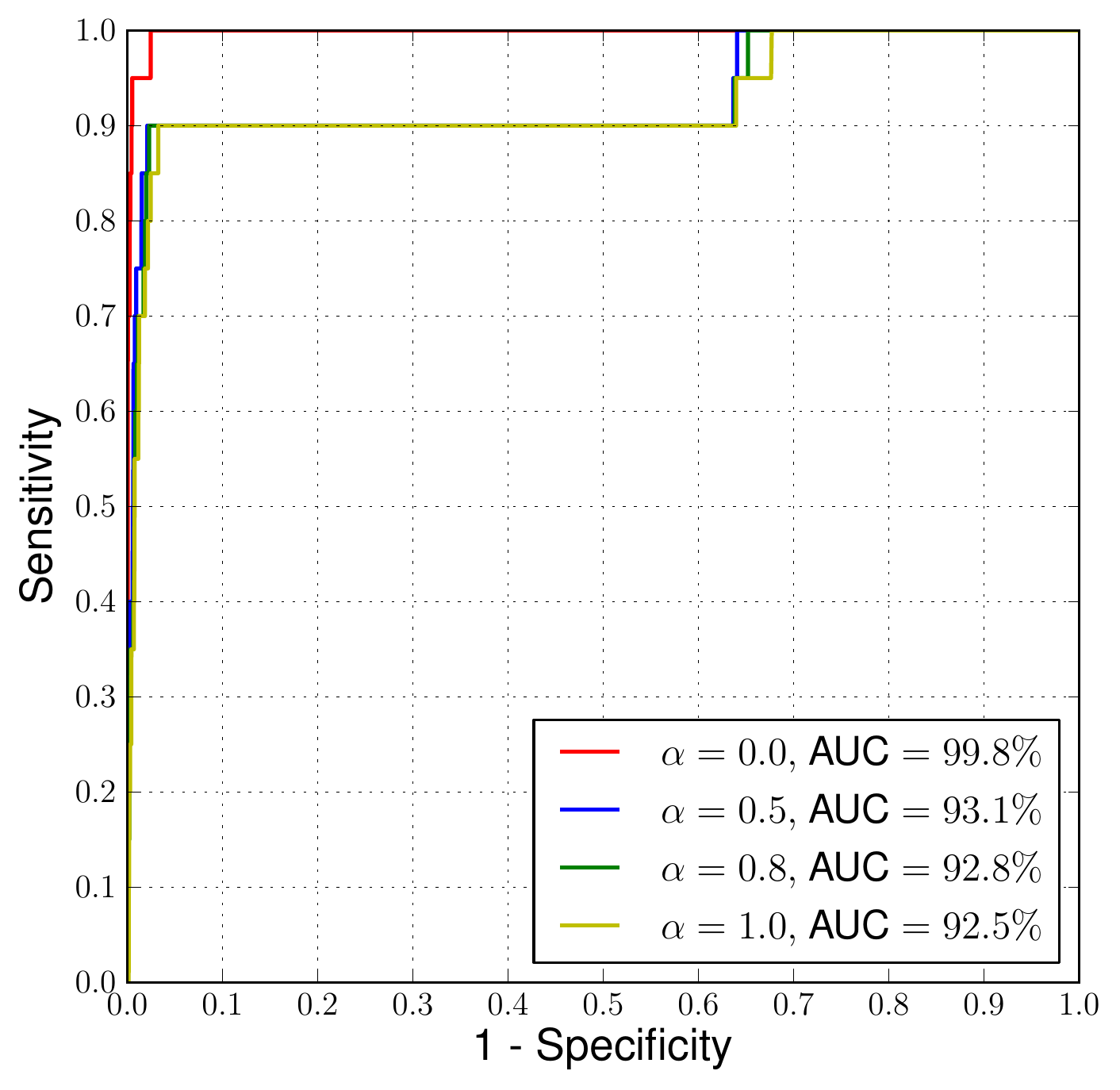}
    \caption{$10-$fold coverage}
\end{subfigure}
\begin{subfigure}{.33\textwidth}
  \centering
  \includegraphics[width=\linewidth]{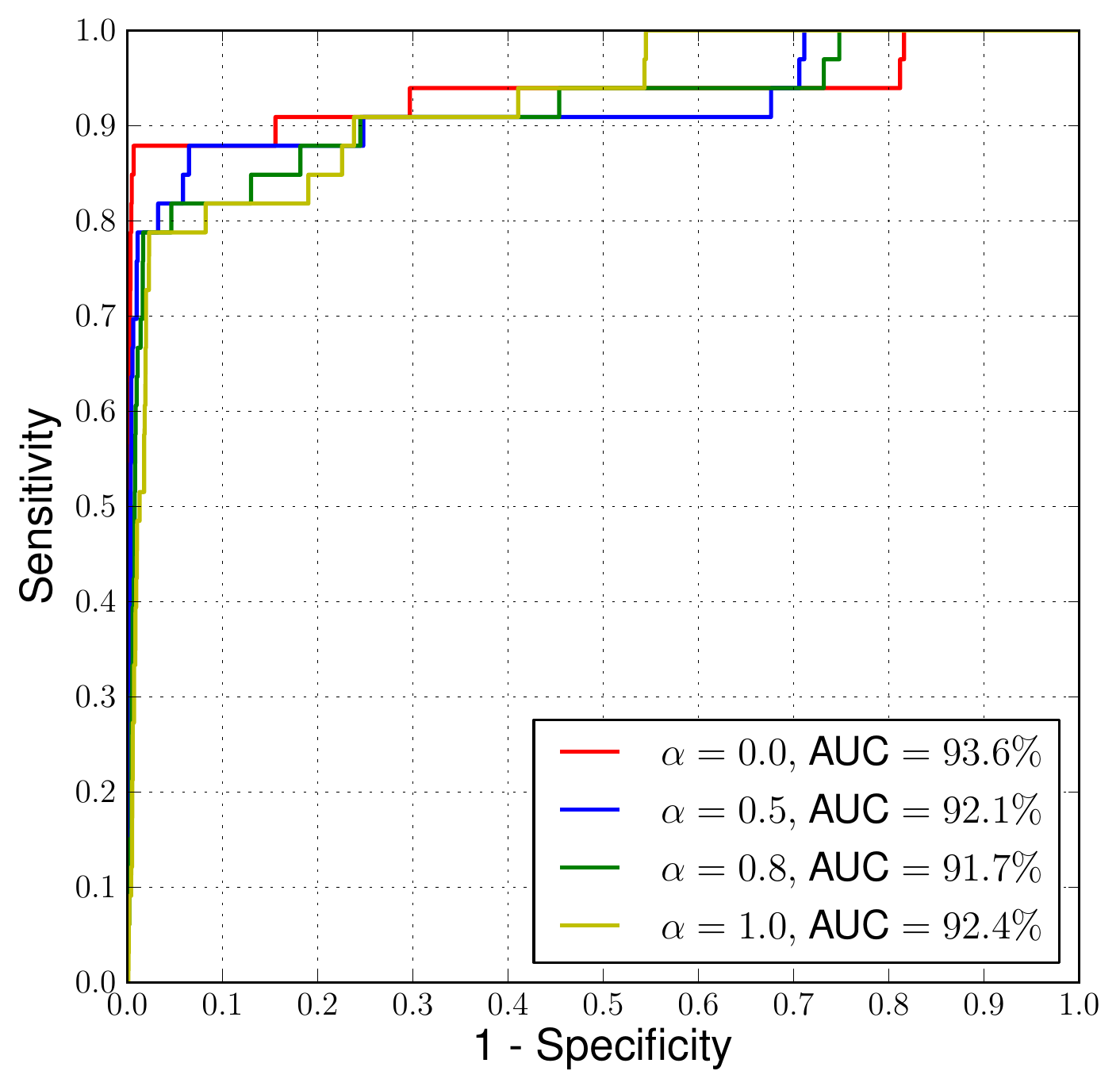}
    \caption{$15-$fold coverage}
\end{subfigure}
\caption{Performance of error-correction over structured positions (i.e. inside a base pair). Subfigures show accuracy when
ART fold parameter is set to 5, 10 and $15-$fold coverage. We also analyze the impact of the parameter $\alpha$ distributing the weights of stacking pair energies vs isostericity scores and use values ranging of $\alpha = \{0, 0.5, 0.8, 1.0\}$. The AUC is indicated in the legend of the figures. Each individual ROC curve represent the average performance over at least 10 experiments.}
\label{fig:16s_bp}
\end{figure}

As in Section~\ref{sec:5S}, we evaluated the predictive power of \RNApyro-computed mutational profiles using a leave-one-out strategy. We picked a sequence at random from the MSA, sequenced/mutated it as above, and run \RNApyro to establish its mutational profile. In this execution, we used values of $\beta=15$ and $\alpha\in [0, 0.5, 0.8, 1.0]$, and set the presumed number of mutations to twice the average error rate made by {\tt ART}, i.e. $4.8\%$, $1.8\%$ and $1.2\%$ for fold coverages of $5$, $10$ and $15$ respectively.
We repeated the whole procedure $12$/$10$/$14$ times for the $5$/$10$/$15-$fold coverage. 
We evaluated the predictive power of the profile by a computing a joint ROC curve for each value of $\alpha$, and each coverage, as described in Section~\ref{sec:5S}. Figure~\ref{fig:16s_all} shows the ROC curves, computed over all positions, while Figure~\ref{fig:16s_bp} only focuses on positions that are paired in the consensus secondary structure.

Our data demonstrates that even on long sequences our algorithm shows interesting potential for 
error-correction applications. First, the AUC values (up to 0.81) when looking at all positions, 
in Fig.~\ref{fig:16s_all},  indicate that a signal has been successfully extracted. When restraining
to structured regions (i.e. base paired regions) we obtain AUC values up to 0.998.  Since 
contributions to the energy and isostericity only arise from structured regions, this was an expected result.

An interesting feature is that best results are almost always when $\alpha=0$, i.e. all the contribution
comes from the isostericity. A notable exception is when the number of mutation is underestimated, 
in that case best performances were observed for non-null values of $\alpha$.

Another observation is when we look at all positions as in Fig.~\ref{fig:16s_all}. The first $20\%$ to $30\%$ of sensitivity are obtained almost without any errors and correspond to the nucleotides in 
structured regions. The rest of the predictions are done on un-paired nucleotides.

\section{Conclusion}
\label{sec:conclusion}

\TODOJerome{Etoffer la conclusion}
In this article we presented a new and efficient way of exploring the mutational landscape of an RNA under structural constraints,
and apply our techniques to identify and fix sequencing errors. In addition, we introduced a new scoring scheme to measure the
likelihood of sequencing errors that combines the classical nearest-neighbour energy model parameters \cite{Turner2010} to the
recently introduced isostericity matrices \cite{Stombaugh2009}. The latter accounting for geometrical discrepancies occurring
during base pair replacements.

We combined our algorithm for exploring the mutational neighbourhood of an input sequence with known secondary structure to
this new pseudo energy model, and create a tool to predict point-wise sequencing errors in structured RNAs. Importantly, our
algorithm runs in  $\Theta(n\cdot(|\Omega|+M^2))$ time and $\Theta(n\cdot(|\Omega|+M))$ memory, where $n$ is the length of
the RNA, $M$ the number of mutations and $\Omega$ the size of the multiple sequence alignment. This achievement enables
us to envision applications to high-throughput sequencing pipe-lines.

We validated our model on the 5s rRNA and 16s rRNA (See Sec.~\ref{sec:results}) and showed that our technology enables us
to recover mutational errors with high accuracy (Area under the ROC curve between $0.95$ and $0.99$ when the mutations are
located in in base paired regions). Interestingly, we observed that using the isostericity matrices alone yields higher performance
than with the nearest-neighbour energy model alone. This finding supports our hypothesis that isostericity matrices provide a
valuable source of informations that can be efficiently used for RNA sequence and structure analysis. Nonetheless, we also found
that the nearest-neighbour model seems to provide a valuable signal when we under-estimate of the number of errors in the input
sequences.

Our techniques are designed to correcting point-wise errors in structured regions (i.e. base paired nucleotides). Nonetheless,
our software \texttt{RNApyro} can be easily combined with other methodologies previously developed for correcting other types
of sequencing errors such as indels in unstructured regions or repeats \cite{Quinlan2008,Quince:2009uq}. In further works, we
also intend to include in our model errors stemming from insertions or deletions. It is indeed theoretically possible to considers
these scenario within our dynamic programming scheme \cite{Waldispuhl:2002fk} with minor impacts of the complexity.

Finally, we hope that integrating our software to current sequencing pipe-lines used in metagenomics studies will permit to
improve the estimate of microbial diversity.

\section{Acknowledgments}
\label{sec:acknowledgments}
The authors would like to thank Rob Knight for his suggestions and comments.
This work was funded by the French Agence Nationale de la Recherche (ANR) through the {\sc Magnum} {\tt ANR 2010 BLAN 0204} 
project (to YP), the FQRNT team grant 232983 and NSERC Discovery grant 219671 (to JW).

\bibliographystyle{jcompbiol}
\bibliography{RNApyro}

\end{document}